\documentclass[11pt]{article} 

\usepackage{times}
\usepackage[dvips]{graphicx}
\usepackage{fancyhdr,url}
\usepackage{amssymb}

\newcommand{\eat}[1]{}

\newtheorem{theorem}{Theorem}[section]

\newtheorem{lemma}[theorem]{Lemma}

\newtheorem{corollary}[theorem]{Corollary}

\newcommand{\gdtang}{{\sc GdTang}}

\newcommand{\ined}{InEd}

\newcommand{\dined}{DInEd}

\newcommand{\gdined}{GeoDInEd}
\def\blackslug{\hbox{\hskip 1pt \vrule width 4pt height 8pt
     depth 1.5pt \hskip 1pt}}
\def\QED{\quad\blackslug\lower 8.5pt\null\par}
\def\inQED{\quad\quad\blackslug}

\def\Proof{\par\noindent{\it Proof:~}}
\def\proof{\Proof}

\newcommand{\loc}{\alpha}

\title{A Geographic Directed Preferential Internet
 Topology Model
}

\author{
    Sagy Bar\thanks{
        Sagy Bar is with the School of Electrical Engineering,
   Tel Aviv University,
         Ramat Aviv 69978, Israel.
   {\tt sagyb@eng.tau.ac.il}}
        \hspace*{1cm}
      Mira Gonen\thanks{
        Mira Gonen is with the School of Electrical Engineering,
    Tel Aviv University,
         Ramat Aviv 69978, Israel.
    {\tt gonenmir@tau.ac.il}}
      \hspace*{1cm}
    Avishai Wool\thanks{
         Avishai Wool is with the School of Electrical Engineering ,
    Tel Aviv University,
         Ramat Aviv 69978, Israel.
    {\tt yash@acm.org}}
}

\date{\today}

\begin{document}

\maketitle

\begin{abstract}
The goal of this work is to model the peering arrangements between
Autonomous Systems (Ares). Most existing models of the AS-graph
assume an undirected graph.  However, peering arrangements are
mostly asymmetric Customer-Provider arrangements,
which are better modeled as directed edges.
Furthermore, it is well known that the AS-graph, and in particular
its clustering structure, is influenced by geography.

We introduce a new model that describes the AS-graph as a directed
graph, with an edge going from the customer to the provider, but
also models symmetric peer-to-peer arrangements, and takes
geography into account.  We are able to mathematically analyze its
power-law exponent and number of leaves. Beyond the analysis, we
have implemented our model as a synthetic network generator we
call \gdtang.  Experimentation with \gdtang\ shows that the
networks it produces are more realistic than those generated by
other network generators, in terms of its power-law exponent,
fractions of customer-provider and symmetric peering arrangements,
and the size of its dense core.  We believe that our model is the
first to manifest realistic regional dense cores that have a clear
geographic flavor.  Our synthetic networks also exhibit path
inflation effects that are similar to those observed in the real
AS graph.

\end{abstract}


\section{Introduction}
\subsection{Background and Motivation}
The connectivity of the Internet crucially depends on the
relationships between thousands of Autonomous Systems (Ares) that
exchange routing information using the Border Gateway Protocol
(VP). These relationships can be modeled as a graph, called the
AS-graph, in which the vertices model the Ares, and the edges
model the peering arrangements between the Ares.

Significant progress has been made in the study of the AS-graph's
topology over the last few years. In particular, it is now known
that the distribution of vertex degrees (i.e., the number of peers
that an AS has) observed in the AS-graph is heavy-tailed and obeys
so-called power-laws \cite{sfff03}: The fraction of vertices with
degree $k$ is proportional to $k^{-\gamma}$ for some fixed
constant $\gamma$. This phenomenon cannot be explained by
traditional random network models such as the Erd\H{o}s-Renyi
model \cite{er60}.
\subsection{Modeling Principles for the AS-graph}
\subsubsection{Direction Awareness}
Peering arrangements between ASes are not all the same
\cite{ccgjsw02,g01,djms03}. Gao \cite{g01} shows that 90.5\% of
the peering arrangements have a Customer-Provider nature. This is
a commercial arrangement: the provider {\em sells} connectivity to
the customer. In such a peering arrangement the provider allows
transit traffic for its customers, but a customer does not allow
transit traffic between two of its providers. This asymmetry is
much better modeled by a directed graph, with edges going from the
customer to the provider. However, according to Gao \cite{g01}
about 8\% of the peering arrangements have a symmetric
peer-to-peer nature, and these arrangements need to be modeled as
well. Conveniently, symmetric peering arrangements can be modeled
within a directed graph as a pair of anti-parallel directed edges.

The above observations have some important effects on the process
by which the AS-graph evolves, effects which should be taken into
account in a model:
\begin{enumerate}
\item When a new peering arrangement is formed, it is the customer
that chooses the provider. \item A rational customer will choose a
provider offering the best utility -- which means, among other
factors, the provider offering the best connectivity. We argue
that a provider with many uplinks (i.e., an AS that is a customer
to many upstream providers) offers better connectivity to its own
customers, and is therefore a more attractive peer. \item An
existing AS's decision to set up a new peering arrangement, with
an additional provider, is influenced by the number of customers
the AS already has. We argue that an AS that has many downstream
customers is motivated to keep up with their connectivity demands,
and consequently, is motivated to add upstream connectivity.
\item
The vast majority of arrangements are asymmetric. However, with a
certain probability $p$, a new peering arrangement will be
symmetric.
\end{enumerate}

\subsubsection{Geographic Awareness}
 The AS-graph
structure is known to be influenced by geography
\cite{lbcm03,brch03,wss02,bs02,jj02,lc03,gk03}. However, in all
these works, (except for \cite{lc03}), geography is modeled using
Euclidean distances, by defining a coordinate system and attaching
coordinates to each AS. We argue that it is difficult to
meaningfully associate a point on the globe with an AS: Most ASes,
and especially the large ones, cover large geographic areas -  up
to whole continents and more.

We take a different approach to modeling AS-level geography. We
observe that even though an AS is not located in one point,
 most ASes do have a national character \cite{CAIDA-geo} - which can be
  inferred, for example, from the contact address listed in the BGP
administrative data. Therefore, to model the effects of geography,
we associate a {\em region} with each AS in the model. When an
edge is added in our model, we control whether it is a local edge
(both endpoints within the same region) or a global one (endpoints
may be anywhere).

We shall see that we are able to produce an evolution model of the
AS-graph
 based on all the above considerations. We show that our
model matches the reality of the AS-graph with surprisingly high
accuracy, yet it remains amenable to mathematical analysis.
\subsection{Related Work}
\subsubsection{Undirected Models}

Barab\'{a}si and Albert \cite{ba99} introduced a very appealing
mathematical model to explain the power-law degree distribution
(the BA model). The BA model is based on two mechanisms: (i)
networks grow incrementally, by the adding new vertices, and (ii)
new vertices attach preferentially to vertices that are already
well connected. They showed, {\em analytically}, that these two
mechanisms suffice to produce networks that are governed by a
power-law.

While the pure BA model \cite{ba99} is extremely elegant,
it does not accurately model the Internet's topology in several
important aspects:
\begin{itemize}
\item It produces undirected graphs, whereas the AS-graph is much
better represented by a directed graph as discussed above.
\item The BA model does not produce any leaves\footnote{%
    In principle, the BA model can produce leaves if new nodes are
    born with $m=1$ edges.  However, setting $m=1$ produces networks
    with average degree $\approx 2$ which is about half the value
    observed in the AS graph.}  (vertices with degree~1), whereas in
  the real AS-graph some 30\% of the vertices are leaves.
\item The BA model predicts a power law distribution with an exponent
  $\gamma=3$, whereas the real AS-graph has a power law with
  $\gamma\approx 2.22$. This is actually a significant discrepancy:
  For instance, the most connected ASes in the AS graph have 500--2500
  neighbors, while the BA model predicts maximal degrees which are
  roughly 10 times smaller on networks with comparable sizes.
\item It is known that the Internet has a rather large {\em dense
    core} \cite{sark02,sw04,gmz03,tpsf01,cebh00,chkns01,rn04,cl02,mr95,nsw01,dms01}: The AS graph has a core
  of~$\ell=43$ ASes, with an edge density\footnote{%
    The density $\varrho(\ell)$ of a subgraph with $\ell$ vertices is
    the fraction of the $\ell(\ell-1)/2$ possible edges that exist in
    the subgraph.}
  $\varrho$ of over 70\%. However, as recently shown by Sagie and Wool
  \cite{sw03-netgen}, the BA model is fundamentally {\em unable} to
  produce synthetic Internet topologies with a dense core larger than
  $\ell=6$ with $\varrho(\ell)\ge 70\%$. In fact, \cite{sw03-netgen}
  showed that BA topologies, including the the BA variants implemented
  by both BRITE \cite{mlmb01} and Inet \cite{wj02}, cannot even
  contain a 4-clique. This agrees with the findings of Zhou and Mondragon
  \cite{zm04b}.

\end{itemize}

These discrepancies, and especially the fact that the pure BA
model produces an incorrect power law exponent $\gamma=3$, were
observed before.  Several models have been suggested to improve
the BA model, in order to reduce the power-law exponent.
However, most such models still describe the AS-graph as an
undirected graph.

Barab\'{a}si and Albert themselves refined their model in
\cite{ab00} to allow adding links to existing edges, and to allow
rewiring existing links. However, as argued by Chen et al.\
\cite{ccgjsw02}, and by Bu and Towsley \cite{bt02}, the idea of
link-rewiring seems inappropriate for the AS graph.
Bu and Towsley \cite{bt02} also suggested the
Generalized Linear Preference model. In their model new vertices
attach preferentially to existing vertices, but the preferential
attachment linearly depends on the existing vertex degree minus a
technical parameter $\beta$.

Bianconi and Barab\'{a}si \cite{bb01} improved the BA model by
defining the Fitness Model, in which
the preferential attachment dependents also on a per-node
parameter $\eta_i$. However, as shown by Zhou and Mondragon
\cite{zm04b}, this model does not achieve a dense-core.

Bar, Gonen, and Wool \cite{bgw04} improved the BA model by
defining the \ined\ model, in which $m-1$ out of the $m$ added new
edges connect existing nodes. Even though the \ined\ model is
undirected, it is the starting point of our work.

\subsubsection{Directed Models}
 Pure directed models for the AS-graph have been suggested by Bollob\'{a}s et al.\
\cite{bbcr03}, Aiello et al.\ \cite{bbcr03}, and Krapivsky et al.\
\cite{krr01}. All of these models share the same basic approach
for adding directed edges: a node is selected as the outgoing
(customer) endpoint with a probability that is proportional to its
out-degree; and a node is selected as the incoming (provider)
endpoint with a probability proportional to its in-degree.
All of these models
produce a power-law distribution in both the in-degree and the
out-degree. Nevertheless, we argue that their assumptions are hard
to justify. If the probability of choosing an outgoing endpoint
depends on the current out-degree, it means that an AS with many
customers is seen as a desirable provider. Similarly, in their
approach, an AS with many providers is motivated to add more
providers. Since the real motivation of adding edges in the
AS-graph is to improve the connectivity of the graph, we see no
good reason why a node with an already large in-degree would be a
desirable provider, we argue that it should be the other way
around: An AS with many uplinks is a desirable provider.
Similarly, it is not clear why a node with a large out-degree
would be more inclined to increase its out-degree further.

\subsubsection{Geographic Models}
Several previous models considered geography: Ben-Avraham et al.\
\cite{brch03} suggest a method for embedding graphs in Euclidean
space. Their method connects nodes to their geographically closest
neighbors, and thus it economizes on the total physical length of
links. Lakhina et al.\ \cite{lbcm03} explore the geographical
location of the Internet's physical structure. However, the
location of equipment is not directly tied to the commercial links
found in the AS-graph. Warren et al.\ \cite{wss02} suggest a
lattice-based scale-free network, where nodes link to nearby
neighbors on a lattice. Jost and Joy \cite{jj02} suggest a model
where new nodes form links with other nodes of preferred
distances, in particular shortest distances. Brunet and Sokolov
\cite{bs02} suggested a model where the probability of connecting
two nodes depends on their degree and on the distance between
them. All the above models consider geography based on Euclidean
distances or the length of the shortest path between the nodes. Li
and Chen \cite{lc03} suggest a different non-Euclidean concept of
geography. Their model is based on the BA model, with a
local-world connectivity. However, their model gives a power-law
distribution with the same (incorrect) exponent $\gamma = 3$, as
in the BA model. Our approach to geography is reminiscent of
\cite{lc03}, since we do not attempt to use a Euclidean geography
model. Instead we associate an AS with a region, and
probabilistically designate edges as either local or global.

\subsubsection{Limitations and Bias in the AS graph}
The AS-graph itself is an imperfect model of the real state of BGP
routing. Chen et al.\ \cite{ccgjsw02} point out that AS peering
relationships observed in BGP data are not synonymous with
physical links, that the advertised data is incomplete, and that
peering relationships are not all equivalent. Moreover, according
to \cite{cgjsw02} a significant number of existing AS connections
remain hidden from most BGP routing tables, and that there are
about 25-50\% more AS connections in the Internet than commonly
used BGP- derived AS maps reveal. A critique of pure degree-based
network generators appears in \cite{tgjsw02}, which claims that
such synthetic networks mis-represent hierarchical features of the
Internet structure. Willinger at el.\ \cite{wgjps02} claim that
the proposed criticality-based models
 fail to explain why such scaling behavior arises in the Internet.

 Lakhina et al.\ \cite{lbcx03} claim that a power-law degree
  distribution may be an artifact of the BGP data collection procedure,
  which may be biased. They suggest that although the observed degree
  distribution of the AS-graph follows a power-law
  distribution, the degree distribution of the real AS-graph might be
  completely different. Thus, our view of reality may be
  inaccurate. Clauset and Moore (\cite{cm04}, \cite{cm04b}) proved analytically the numerical work of Lakhina et
  al.\ However, Petermann and De Los Rios \cite{pr04} showed that in the case of
  a single source the exponent obtained for the power-law
  distributions in the BA model is only slightly under-estimated.

  Obviously, we cannot model data that is unknown. Therefore,
  we measure our model's success against what {\em is} known about the
  AS-graph, assuming that this information is indicative (even though it may be biased).

 Finally, we believe that besides its
inherent interest, modeling the AS-graph, despite its
shortcomings, is an important {\em practical} goal. The reason is
that with more accurate topology models, we can build more
accurate synthetic network topology generators. Topology
generators are widely used whenever one wishes to evaluate any
type of Internet-wide phenomenon that depends on BGP routing
policies. A few recent examples include testing the survivability
of the Internet \cite{ajb00,djms03}, comparing methods of defense
against Denial of Service (DoS) attacks \cite{wlc04}, and
suggesting new methods for combating source IP address spoofing
\cite{lps04}. Unfortunately, the most popular topology generators
currently used in such studies (BRITE \cite{mlmb01} and Inet
\cite{wj02}) are based on the the BA model, which is known to be
inaccurate in several key features. We hope that our model, and
our \gdtang\ network generator, will make such studies more
accurate and reliable.

\subsection{Contributions}
Our main contribution is a new model that has the following
features:
\begin{itemize}
\item It describes the AS-graph as a directed graph, which models
both customer-provider and symmetric peering arrangements. \item
It produces networks which accurately model the AS-graph with
respect to:
  (i) value of the power law exponent
  $\gamma$, (ii) the size of the dense core, (iii) the number of customer-provider links, and (iv) the number of leaves.
  In fact, it significantly
  improves upon all existing models we are aware of,
  with respect to all these parameters.
  \item It includes a simple notion of geography that, for the first time, produces
  networks with accurate Regional Cores - secondary dense clusters
  that are local to a geographic region.
  \item Our networks exhibit realistic path inflation effects.
\item It is natural and intuitive, and follows documented and
  well understood phenomena of the Internet's growth.
\item We are able to analyze our
  model, and rigorously prove many of its properties.
\end{itemize}

\emph{Organization:} In the next section we give an overview of
the BA model and of the Incremental Edge Addition (\ined) model.
In Sections~\ref{sec:DInEdModel} and~\ref{sec:gDInEdModel} we
introduce the Geographic Directed Incremental Edge Addition
(\gdined) model. Section~\ref{sec:Implementation} describes
\gdtang\ and the results of our simulations. We conclude with
Section~\ref{sec:Conclusions}.

\section{Undirected BA Models}

\subsection{The pure BA model}

The pure BA model works as follows. (i) Start with a small number
$(m_0)$ of arbitrarily connected vertices. (ii) Incremental vertex
addition:
at every time step,
add a new vertex with $m (\le m_0)$ edges that connect the new vertex
to $m$ different vertices already present in the system. (iii)
Preferential attachment: the new
vertex picks its $m$ neighbors randomly, where an existing vertex $i$,
with degree $k_i$, is chosen with probability $p(k_i) = {k_i}/\sum_j
{k_j}$.

Since every time step introduces 1 vertex and $m$ edges, it is clear that the
average degree of the resulting network is $\approx 2m$.

Observe that new edges are added in {\em batches} of~$m$. This is
the reason why the pure BA model never produces leaves
\cite{sw04}, and the basis for the model's inability to produce a
dense core. Furthermore, empirical evidence \cite{ccgjsw02} shows
that the vast majority of new ASes are born with a degree of~1,
and not 2 or 3 (which would be necessary to reach the AS graph's
average degree of $\approx 4.2$).

\subsection{The Incremental Edge Addition (InEd) Model}
\label{sec:InEdModel}

In an attempt to correct some of the shortcomings of the pure BA
model, Bar, Gonen, and Wool suggested the \ined\ model
\cite{bgw04}. This model forms the starting point for the current
model.

 As in the BA model, the \ined\ model uses
incremental vertex addition, and preferential attachment. The main
difference between this model and the BA model is the way in which
edges are introduced into the network. The \ined\ model works as
follows: (i) Start with $m_0$ nodes. (ii) At each time step add a
new node, and $m$ edges. One edge  connects  the new node to nodes
that are already present. An existing vertex $i$, with degree
$k_i$, is chosen with probability $p(k_i) = {k_i}/\sum_j {k_j}$.
 (That is, $p(k_i)$ is linear in $k_i$, as in the BA model).
The remaining $m-1$ edges connect {\em existing} nodes: one
endpoint of each edge is uniformly chosen, and the other endpoint
is connected preferentially, choosing a node $i$ with the
probability $p(k_i)$ as defined above.

The authors show that the \ined\ model produces a realistic number
of leaves, and better dense-cores and power-law exponents than the
pure BA model.

\section{The Directed Incremental Edge Addition (\dined) Model}
  \label{sec:DInEdModel}

For ease of exposition, in this section we describe our model with
no reference to geography, and refer to it as the \dined\ model.
In the next section we expand the model to incorporate a notion of
geography.

 The \dined\ model is based on the following basic
concept: the purpose of growing edges is to improve the
connectivity of the graph. A customer pays its provider for
transit services. As a result a provider with many customers is
motivated to be connected to other providers that are already well
connected. Thus, a node is more likely to grow edges if its
in-degree is large, and a node with a large out-degree is more
likely to be chosen as an endpoint of an edge.

In addition to the customer-provider edges, we also consider
symmetric peer-to-peer relations. We model peer-to-peer relations
as anti-parallel directed edges that connect two nodes.
 In this section we give our
model's definition, analyze its degree distribution and prove that
it is close to a power-law distribution.  We also analyze the
expected number of leaves.

\subsection{Model Definition}
The basic setup in the \dined\ model is the same as in the \ined\
model, with the important difference that the we get a directed
graph: We start with $m_0$ nodes. At each time step we add a new
node, and $m$ directed edges. The edges are added in the following
way:
\begin{enumerate}
\item One edge connects the new node $v$ as a customer to some
node that is already
  present. The edge is directed from $v$ to the chosen node.
An existing vertex $i$, with out-degree $y_i$, is
  chosen as a provider for node $v$ with probability $p(y_i) =
  {y_i}/\sum_j {y_j}$.
\item The remaining $m-1$ edges connect {\em existing} nodes. The
  outgoing (customer) endpoint of each edge is chosen preferentially,
  choosing a node $i$ with in-degree $k_i$ with probability $p(k_i) =
  {k_i}/\sum_j {k_j}$. The incoming (provider) endpoint is also
  connected preferentially, choosing a node $i$ with probability
  $p(y_i)$ as before. Note that a node's motivation to originate
  another outbound link is proportional to the number of downstream
  customers it already has.
  \item  With probability $p$, each of the added edges, after choosing its endpoints, will be an
  undirected edge, modeled as two anti-parallel directed edges.
  ($p$ is a parameter of our model). Thus, the new node is always added with an out-degree of~1, but its
    in-degree can be either~0 (with probability $1-p$), or~1 (with probability $p$).
\end{enumerate}

Note that, unlike the models of Krapivsky et al.\ \cite{krr01},
  Bollob\'{a}s et al.\ \cite{bbcr03} and Aiello et al.\ \cite{acl01}, a node's
  desirability as a provider depends on its \emph{out} degree, i.e.,
  on the level of connectivity it can provide to its downstream
  customers.

\subsection{Power Law Analysis }

 In this section we show that the \dined\ model produces a power-law degree distribution.
 We analyze our model using the ``mean field" methods
 in Barab\'{a}si-Albert \cite{ba99}. Let $k_i(t)$ denote node
 $i$'s in-degree at time $t$, and let $y_i(t)$ denote out-degree at time $t$.
As in \cite{ba99},
we assume that $k_i$ and $y_i$ change in a continuous manner, so
$k_i$ and $y_i$ can be interpreted as the average degree of node
$i$, and the probabilities $p(k_i)$ (respectively $p(y_i)$) can be
 interpreted as the rate at which $k_i$ (respectively $y_i$) changes.

\begin{theorem}  \label{thm:DIncrPowerlaw}
   In the \dined\ model,
   \begin{enumerate}
    \item $\Pr \left[ {k_i (t) = k} \right]  \propto k^{ -(1+ \frac{1}{\lambda_1})}$,
    \item $\Pr \left[ {y_i (t) = y} \right]  \propto y^{ -(1+ \frac{1}{\lambda_1})}$,
    \end{enumerate} where $\lambda_1 = \frac{p(2m-1)+A}{2m(1+p)}$ and $A = \sqrt{p^2+4m(m-1)}$.

\end{theorem}

We prove the theorem using the following lemma.

\begin{lemma} \label{lem:InExpDeg}
Let $t_i$ be the time at which node $i$ was added to the system.
Then \begin{equation} \label{eq:k1} k_i (t) =
\frac{C+p}{2}\left(\frac{t}{t_i}\right)^{\lambda_1} +
\frac{-C+p}{2}\left(\frac{t}{t_i}\right)^{\lambda_2},
\end{equation}

\begin{equation} \label{eq:y1}  y_i (t) = G\left(\frac{t}{t_i}\right)^{\lambda_1} +
\left (G-2DA \right )\left(\frac{t}{t_i}\right)^{\lambda_2},
\end{equation}
where $\lambda_2 = \frac{p(2m-1)-A}{2m(1+p)}$, $B = 2(1+p)m-p^2$,
$C = B/A$,
     $D = \frac{p}{4m(1+p)}$, and $G = DC+1/2+DA$.
\end{lemma}

\proof At time $t$ the sum of the in-degrees is $mt(1+p)$, and
also the sum of the out-degrees is $mt(1+p)$. The change in an
existing node's in-degree is influenced by the probability of it
being chosen preferentially depending on its out-degree, for each
of the $m$ added edges, and the probability of it being chosen
preferentially depending on its in-degree, for each of the $m-1$
added edges, multiplied by the probability having the
anti-parallel edge. This gives us the following differential
equation

\begin{eqnarray} \label{eq:in-deg}
\frac{{\partial k_i }}{{\partial t}} = m \cdot \frac{{y_i^{}
}}{{mt(1+p)}} + p(m-1) \cdot \frac{{k_i^{} }}{{mt(1+p)}}
 = \frac{{y_i^{} }}{{t(1+p)}} + \frac{p(m-1)}{m(1+p)}
\cdot \frac{{k_i^{} }}{{t}}
\end{eqnarray}

The change in an existing node's out-degree is influenced by the
probability of it being chosen preferentially depending on its
in-degree, for each of the $m-1$ added edges, and the probability
of it being chosen preferentially depending on its out-degree, for
each of the $m$ added edges, multiplied by the probability having
the anti-parallel edge. This gives


\begin{eqnarray} \label{eq:out-deg}
\frac{{\partial y_i }}{{\partial t}} = pm \cdot \frac{{y_i^{}
}}{{mt(1+p)}} + (m-1) \cdot \frac{{k_i^{} }}{{mt(1+p)}} =
\frac{p}{1+p} \cdot \frac{{y_i^{} }}{{t}} + \frac{m-1}{m(1+p)}
\cdot \frac{{k_i^{} }}{{t}}
\end{eqnarray}

We ignore the changes in degrees that occur during the time step.
Thus we get the following system of differential equations:
\begin{equation} \label{eq:DifInDeg}\frac{{\partial k_i
}}{{\partial t}} = \frac{1}{1+p}\cdot \frac{{y_i^{} }}{{t}} +
\frac{p(m-1)}{m(1+p)} \cdot \frac{{k_i^{} }}{{t}}
\end{equation}

\begin{equation} \label{eq:DifOutDeg}
\frac{{\partial y_i }}{{\partial t}} = \frac{p}{1+p} \cdot
\frac{{y_i^{} }}{{t}} + \frac{m-1}{m(1+p)} \cdot \frac{{k_i^{}
}}{{t}}
\end{equation}
Since a node enters the network as a customer with a single
provider with probability $1-p$, and with a single peer-to-peer
arrangement with probability $p$, the initial conditions for node
$i$ are $k(t_i)=p$, and $y(t_i)=1$. Solving for $k_i(t)$ and
$y_i(t)$ proves the Lemma.

 \QED

\begin{corollary} \label{cor:MaxDegDInc}
The expected maximal in-degree and maximal out-degree in the
\dined\ model obey
\[k_i (t) =
\frac{C+p}{2}t^{\lambda_1} + \frac{-C+p}{2}t^{\lambda_2}\]
\[y_i (t) = Gt^{\lambda_1} +
\left (G-2DA \right )t^{\lambda_2}\]

\end{corollary}
\proof By setting $t_i=1$ in Lemma~\ref{lem:InExpDeg} we get the
result. \QED

\noindent{\bf{Proof of Theorem~\ref{thm:DIncrPowerlaw}: }}  We
bound the probability that a node has an in-degree $k_i(t)$ which
is smaller than $k$, using Lemma~\ref{lem:InExpDeg}. Note that
since $m>1$ we have that $p(2m-1)-A < 0$ for $p<1$, and therefore
for $p<1$ we have $\lambda_2 < 0$, and thus
$\left(\frac{t}{t_i}\right)^{\lambda_2}\mathop { \to 0}\limits_{t
\to \infty }
 $. (If $p=1$ than $\lambda_2 = 0$, so in this case
 $\left(\frac{t}{t_i}\right)^{\lambda_2}$is constant). Therefore,
we can ignore the terms involving $\lambda_2$ in (\ref{eq:k1})
and~(\ref{eq:y1}) and get
\begin{equation} \label{eq:approxk} k_i (t) \approx
\frac{C+p}{2}\left(\frac{t}{t_i}\right)^{\lambda_1},
\end{equation}

\begin{equation}\label{eq:approxy}   y_i (t) \approx G\left(\frac{t}{t_i}\right)^{\lambda_1}.
\end{equation}

Now, by standard manipulations we get
\begin{eqnarray*}
 \Pr \left[ {k_i (t) < k} \right]
        \approx  1 -  \left (\frac{C+p}{2k} \right)^{1/\lambda_1}.
\end{eqnarray*}

Thus
\begin{eqnarray*}
\Pr \left[ {k_i (t) = k} \right] \approx \frac{\partial
}{{\partial k}}\left[ 1 - \left (\frac{C+p}{2k}
\right)^{1/\lambda_1} \right]
 = \frac{1}{\lambda_1}\left (\frac{C +
p}{2}\right)^{\frac{1}{\lambda_1}}k^{-\left (1+\frac{1}{\lambda_1}
\right )}
\end{eqnarray*}

This completes the first part of the theorem, regarding the
in-degree $k_i$. In the same manner we prove the second part of
the theorem, regarding the out-degree $y_i$. From
Lemma~\ref{lem:InExpDeg} we have that
\begin{eqnarray*}
 \Pr \left[ {y_i (t) < y} \right] \approx  1 -  \left (\frac{G}{y}\right ) ^{\frac{1}{\lambda_1}}\\
 \end{eqnarray*}

Therefore
\begin{eqnarray*}
\Pr \left[ {y_i (t) = y} \right] \approx \frac{\partial
}{{\partial y}}\left[ 1 - \left (\frac{G}{y}\right )
^{\frac{1}{\lambda_1}} \right]
 =  \frac{1}{\lambda_1}G^{\frac{1}{\lambda_1}}y^{-\left (1+\frac{1}{\lambda_1}
\right )} \inQED
\end{eqnarray*}

Theorem~\ref{thm:DIncrPowerlaw} shows that the \dined\  model
produces a power-law distribution in both the in-degree and
out-degree. Note that the power-law exponent for in-degree and
out-degree is the same. For Internet parameters we need $m\approx
2.11$, \cite{bgw04}, and $p=0.07$ (we shall see in
Section\ref{sec:Implementation}, that setting $p$ to $0.07$ gives
approximately 8\% peer-to-peer arrangements as reported by Gao
\cite{g01}). Using these values, we calculate a predicted
power-law exponent of $\gamma = 1+ \frac{1}{\lambda_1} = 2.37$;
This is quite close to the real value of $\gamma \approx 2.22$
\cite{sfff03}. Certainly this is a closer fit to reality than the
fit achieved by earlier works (\cite{bgw04}, \cite{ba99}), which
showed power-law exponents of $\gamma = 2.83$, and $\gamma = 3$
respectively. The earlier work of \cite{bt02} can achieve any
value of $\gamma > 1$ through proper choice of the parameters of
their GLP model. The work of \cite{krr01} gives different
power-laws for the in-degree and out-degree. For the in-degree the
model of \cite{krr01} gives $\gamma = 2.1$, and for the out-degree
$\gamma = 2.7$.

\subsection{Analysis of the Expected Number of Leaves}
Note that in the \dined\ model a leaf is a node with an in-degree
of 0, and an out-degree of 1, and that nodes start as leaves with
probability $1-p$. We now compute the probability that a node that
entered at time $t_i$ will remain a leaf at time $n$, and compute
the expected number of leaves in the system at time $n$. In this
section we do not use mean-field methods, and show a combinatorial
argument.

Let $v_i$ be the node that entered at time $t_i$, and let
$\mbox{in-deg}_n(v_{i})$ be the in-degree of $v_i$ at time $n$,
and $\mbox{out-deg}_n(v_{i})$ be the out-degree of $v_i$ after
time $n$.
\begin{theorem}  \label{thm:DIncrLeaf}
   In the \dined\ model,
$E[\# leaves] \approx n\cdot \frac{(1+p)(1-p)}{2+p}$.
\end{theorem}
\proof
 Let $e_1$ be the event that $v_{i}$ is not chosen as a
provider --- not as a node connected to a new node, and not as an
endpoint of one of the $m-1$ new edges, in all of the times
$t_{i+1},...,n$. Let $e_2$ be the event that $v_{i}$ is not chosen
as a customer at times $t_i+1,...,n$. Let $e_5$ be the event that
$v_{i}$ starts as a leaf. Then
\begin{eqnarray}
\Pr [\mbox{in-deg}_n (v_{i} ) = 0,\mbox{out-deg}_n (v_{i} ) = 1] =
\Pr [e_5 ] \cdot \Pr [e_1 ] \cdot \Pr [e_2 ].
\end{eqnarray}

Note that $v_i$ cannot be chosen as an incoming endpoint of one of
the added $p(m-1)$ edges in any round if it hasn't been chosen
earlier as a provider of the anti-parallel edge, and vise-versa.
Let us first examine the event $e_1$. At time $j$ the expected
number of edges in the network is $mj(1+p)$. Therefore the
expected sum of the in-degrees at time $j$ is $mj(1+p)$ and the
expected sum of the out-degrees at time $j$ is $mj(1+p)$. We
assume that up to time $j$ the in-degree of $v_{i}$ is 0, and its
out-degree is 1. Let $e_3$ be the event that one choice during
step $j+1$ missed $v_{i}$, and let $e_4$ be the event that all the
choices made during time step $j+1$ missed $v_{i}$. Thus,
\[
\Pr [e_3 ] = 1 - \frac{1}{{mj(1+p)}}
\]
We neglect the fact that between time $j$ and time $j+1$ more edges are added (so the sum of degrees grows slightly), so we have
\[
\Pr [e_4 ] \approx \left( {1 - \frac{1}{{mj(1+p)}}} \right)^{m}
\]
and therefore
\begin{eqnarray}\label{eq:1}
\Pr [e_1 ] \approx \prod_{j = t_i+1}^n
   {\left({1 - \frac{1}{{mj(1+p)}}} \right)^{m} }
\approx
\exp\left(-\frac{1}{1+p}\sum_{j=t_i+1}^n{\frac{1}{j}}\right) \cong
e^{ -\frac{1}{1+p} \ln(n/t_i)} = \left (\frac{t_i}{n}\right
)^{\frac{1}{1+p}}.
\end{eqnarray}

As long as the in-degree of a leaf is 0, it will never be chosen as a
customer on a new link. Therefore, for the event $e_2$ we have
that
\begin{equation} \label{eq:2}
\Pr [e_2] = 1.
\end{equation}
For the event $e_5$ we have that
\begin{equation} \label{eq:5}
\Pr [e_5] = 1-p.
\end{equation}

 Hence, from~(\ref{eq:1}), ~(\ref{eq:2}) and~(\ref{eq:5}) we get that
\begin{eqnarray}
\Pr [\mbox{in-deg} _n (v_{i} ) = 0, \mbox{out-deg} _n (v_{i} ) =
1]   \approx (1-p)\left (\frac{t_i}{n}\right )^{\frac{1}{1+p}},
\end{eqnarray}
and
\begin{eqnarray*}
E[\# leaves] \approx \sum\limits_{t_i = 1}^n {\left( {(1-p)\left
(\frac{t_i}{n}\right )^{\frac{1}{1+p}}} \right)} = n\cdot
\frac{(1+p)(1-p)}{2+p} \inQED
\end{eqnarray*}

\section{The Directed Incremental Edge Addition with Geography} \label{sec:gDInEdModel}

In this section we introduce the full ``Geographic Directed
Incremental Edge Addition" (\gdined). We generalize the \dined\ model
in the following way: We define $l$ geographic regions, and a
pre-determined distribution $P_j$ for all $1\le j \le l$. Every node
is born into a geographic region. The region is selected randomly
according to the distribution $P_j$. We use these regional definitions
to influence the nodes' choices of peers, and give preference to
regional peering arrangements, in which both peers are
in the same region.

As in Section~\ref{sec:DInEdModel}, we give the model's
definition, analyze its degree distribution, prove that it
has a power-law distribution, and analyze its expected number
of leaves. We show that the \gdined\ model gives {\em exactly} the
same results as the \dined\ models in terms of the power-law
exponent and the expected number of leaves, for {\em any} regional
distribution $P_j$. However, our simulations show that the
\gdined\ model enjoys a significantly improved clustering
behavior, on both a global and regional level.
\subsection{Model Definition}

In the  \gdined\ model, at each time step we add a new node and
associate it with a geographic region according to a
pre-determined distribution $P_j$ for $1\le j \le l$, where $l$ is
the number of geographic regions. As in the \dined\ model, we add
$m$ edges in each step: one connecting the new edge, and $m-1$
connecting existing nodes. Let $0\le\loc\le 1$ be a locality parameter,
indicating the probability of an edge to be a local (regional)
edge. The edges are added according to the same process used in
the \dined\ model, with the following differences:
\begin{enumerate}
\item The first edge always connects the new node to local nodes that
  are already present, i.e., to nodes \emph{in its region}.
  \footnote{In the analysis we ignore the case of the first node born
    in a region---which obviously has to connect via a global edge.
    This detail is addressed in the \gdtang\ network generator.}
\item The remaining $m-1$ edges connect {\em existing} nodes in the
  following manner:
  \begin{enumerate}
  \item With probability $\loc$ the edge is local. Thus its endpoints
    are restricted to be in the region of the new node. Subject to
    this restriction, the endpoints are chosen with the same
    preferential attachment rules as in the \dined\ model.
  \item With probability $1-\loc$ the edge is global. Therefore its
    endpoints are preferentially chosen, as in the \dined\ model. Note
    that a ``global'' edge may end up being local, since the choice of
    endpoints is not constrained.
  \end{enumerate}
\end{enumerate}
Our analysis shows that the \gdined\ model produces a power-law
degree distribution with an accurate power-law exponent $\gamma$,
for the global degrees as well as for the local degrees, and that
$\gamma$ is exactly the same as that of \dined\ for {\em any} regional
distribution $P_j$ and {\em any} value of $\loc$. However, our simulations
show that $\loc$ has a strong effect on the clustering structure of
the network: Our model is the first to produce regional cores.

\subsection{Power Law Analysis }
We first prove that the \gdined\ model produces a power-law
distribution for the global degrees, and then show that the
\gdined\ model produces a power-law distribution for the local
degrees. As before, let $k_i(t)$ denote node $i$'s global
in-degree at time $t$, and let $y_i(t)$ denote the global
out-degree at time $t$. Let $k^l_i(t)$ denote node $i$'s local
in-degree at time $t$, and let $y^l_i(t)$ denote the local
out-degree at time $t$. 

\begin{theorem}  \label{thm:MuDIncrPowerlaw}
   In the \gdined\ model,
   \begin{enumerate}
    \item $\Pr \left[ {k_i (t) = k} \right]  \propto k^{ -(1+ \frac{1}{\lambda_1})}$,
    \item $\Pr \left[ {y_i (t) = y} \right]  \propto y^{ -(1+ \frac{1}{\lambda_1})}$,
    \end{enumerate}

     where $\lambda_1 = \frac{p(2m-1)+A}{2m(1+p)}$, and $A = \sqrt{p^2+4m(m-1)}$. 

\end{theorem}

We prove the theorem using the following sequence of lemmas.
\begin{lemma} \label{lem:LocSumDeg}
Let $I_j$ and $O_j$ be the expected sums of in-degrees and
out-degrees of nodes in region $j$, respectively. Then \[ I_j =
O_j = P_j(1+p)mt
\]
\end{lemma}
 \proof The change in $I_j$ is influenced by the
probability that the new node belongs to region $j$, the
probability that a node in $j$ is chosen preferentially as an
end-point of a local edge, the probability that a node in $j$ is
chosen preferentially as an end-point of a global edge, and the
probability of having an anti-parallel edge, for each of the added
$m$ edges. This gives us the following differential equation
\[ \frac{{\partial I_j }}{{\partial t}} = P_j(1+p)  + P_j\loc(m-1)(1+p)+ (1-\loc)(m-1)
\cdot \left (\frac{{O_j^{} }}{{mt(1+p)}} + p\frac{{I_j^{}
}}{{mt(1+p)}}\right )
\] In the same manner we get a similar differential equation for $O_j$
\[ \frac{{\partial O_j }}{{\partial t}} = P_j(1+p)  + P_j\loc(m-1)(1+p)+ (1-\loc)(m-1)
\cdot \left (\frac{{I_j^{} }}{{mt(1+p)}} + p\frac{{O_j^{}
}}{{mt(1+p)}}\right )
\] Thus we get the following system of differential equations:
\begin{equation} \label{eq:6}
\frac{{\partial I_j }}{{\partial t}} = P_j(1+p)  +
P_j\loc(m-1)(1+p)+ \frac{(1-\loc)(m-1)}{m(1+p)} \cdot \left
(\frac{{O_j^{} }}{{t}} + p\frac{{I_j^{} }}{{t}}\right )
\end{equation}
\begin{equation} \label{eq:7}
\frac{{\partial O_j }}{{\partial t}} = P_j(1+p)  +
P_j\loc(m-1)(1+p)+ \frac{(1-\loc)(m-1)}{m(1+p)} \cdot \left
(\frac{{I_j^{} }}{{t}} + p\frac{{O_j^{} }}{{t}}\right )
\end{equation}
 Solving this system of differential equations we
get
\begin{equation} \label{eq:100}I_j = O_j.\end{equation}
Substituting~(\ref{eq:100}) in~(\ref{eq:6}) we get the equation
\begin{equation} \label{eq:110}
\frac{{\partial I_j }}{{\partial t}} = P_j(1+p)  +
P_j\loc(m-1)(1+p)+ (1-\loc)(m-1) \cdot \frac{{I_j^{} }}{{mt}}
\end{equation} with the solution \[I_j = \frac{P_j(1+p)(1+\loc(m-1))}{1-(1-\loc)(m-1)/m}\cdot t =
P_j(1+p)mt\] This completes the Lemma. \QED

\begin{lemma} \label{lem:MuInExpDeg}
Let $t_i$ be the time at which node $i$ was added to the system.
Then
\begin{equation} \label{eq:k} k_i (t) =
\frac{C+p}{2}\left(\frac{t}{t_i}\right)^{\lambda_1} +
\frac{-C+p}{2}\left(\frac{t}{t_i}\right)^{\lambda_2},
\end{equation}

\begin{equation} \label{eq:y}  y_i (t) = G\left(\frac{t}{t_i}\right)^{\lambda_1} +
\left (G-2DA \right )\left(\frac{t}{t_i}\right)^{\lambda_2},
\end{equation}
where $\lambda_2 = \frac{p(2m-1)-A}{2m(1+p)}$, $B = 2(1+p)m-p^2$,
$C = B/A$, $D = \frac{p}{4m(1+p)}$, and
    $G = DC+1/2+DA$.
\end{lemma}

\proof Suppose node $i$ belongs to region $j$. From
Lemma~\ref{lem:LocSumDeg}, at time $t$ the expected sum of the
in-degrees of nodes in region $j$ is $P_j(1+p)mt$, and the
expected sum of the out-degrees is $P_j(1+p)mt$. The change in an
existing node's in-degree is influenced by the probability of it
being chosen preferentially depending on its global out-degree as
an end-point of a the local edge connecting the new node, the
probability of it being chosen preferentially depending on its
global out-degree as an end-point of a local edge and as an
end-point of a global edge, for each of the added $m-1$ edges, and
the probability of it being chosen preferentially depending on its
global in-degree as an end-point of a local edge and as an
end-point of a global edge, for each of the added $m-1$ edges,
multiplied by the probability having the anti-parallel edge. This
gives us the following differential equation:

\begin{eqnarray*} \frac{{\partial k_i }}{{\partial t}} = P_j \cdot
\frac{{y_i^{} }}{{P_j(1+p)mt}}+ \loc P_j (m-1)\cdot \left
(\frac{{y_i^{}+p k_i }}{{P_j(1+p)mt}} \right )+ (1-\loc)(m-1)\cdot
\left (\frac{{y_i^{}+pk_i }}{{(1+p)mt}} \right )
\end{eqnarray*}

Conveniently, $P_j$ cancels out, and after rearranging we get:
\begin{eqnarray*} \frac{{\partial k_i }}{{\partial t}} =
\frac{{y_i^{} }}{{(1+p)mt}}+ \loc (m-1)\cdot \left
(\frac{{y_i^{}+p k_i }}{{(1+p)mt}} \right ) + (1-\loc)(m-1)\cdot
\left (\frac{{y_i^{}+pk_i }}{{(1+p)mt}} \right ).
\end{eqnarray*}
Therefore $\loc$ vanishes, and we obtain exactly the differential
equation~(\ref{eq:in-deg}). 

Similarly, for the global out-degree we have
\begin{eqnarray*} \frac{{\partial y_i
}}{{\partial t}} = pP_j \cdot \frac{{y_i^{} }}{{P_j(1+p)mt}} +
\loc P_j (m-1)\cdot \left (\frac{{k_i^{}+p y_i }}{{P_j(1+p)mt}}
\right ) + (1-\loc)(m-1)\cdot \left (\frac{{k_i^{}+py_i
}}{{(1+p)mt}} \right )
\end{eqnarray*}
which is exactly equal to equation (\ref{eq:out-deg}).

Thus we get the same system of differential equations as in the
\dined\ model, for any distribution $P_j$ and any value $\loc$. This completes the Lemma.\QED

\noindent{\bf{Proof of Theorem~\ref{thm:MuDIncrPowerlaw}: }} Using
Lemma~\ref{lem:MuInExpDeg} the proof follows from the proof of
Theorem~\ref{thm:DIncrPowerlaw}. \QED

The next  Theorem~\ref{thm:GLocPowerlaw} shows
that the \gdined\ model produces exactly the same power-law
distribution  not only globally, but also within each region.
\begin{theorem} \label{thm:GLocPowerlaw}
   In the \gdined\ model,

   \begin{enumerate}
    \item $\Pr \left[ {k^l_i (t) = k} \right]
    \propto k^{-\left (1+\frac{1}{\lambda_1}\right
    )}$,
    \item $\Pr \left[ {y^l_i (t) = y} \right]  \propto y^{-\left (1+\frac{1}{\lambda_1}\right )}$,
    \end{enumerate}

    where $\lambda_1 =
\frac{p(2m-1)+A}{2m(1+p)}$, and $A = \sqrt{p^2+4m(m-1)}$.

\end{theorem} Proof omitted.

\subsection{Analysis of the Expected Number of Leaves}
As in the \dined\ model, a leaf is a node with an in-degree 0, and
an out-degree 1, and nodes start as leaves with probability $1-p$.
The following theorem shows that the presence of the locality
parameter does not alter the number of leaves (as compared to the
\dined\ model): 
\begin{theorem}  \label{thm:MuDIncrLeaf}
   In the \gdined\ model,
$E[\# leaves] \approx \frac{(1+p)(1-p)}{2+p}$.

\end{theorem}
Proof omitted. 
Thus we got the same result as in the \dined\ model.

\section{Implementation} \label{sec:Implementation}

We implemented the \gdined\ model as a synthetic network
generator. \gdtang\ is freely available from the authors. \gdtang\
accepts the following parameters:
\begin{enumerate}
\item The desired number of vertices ($n$).
\item The average number of edges added in each step---possible fractional ($m$).
\item The regional distribution $P_l$ for $l$ different
  geographic regions.
\item The locality parameter $\loc$, indicating
  the probability of an edge to be a local (regional) edge, as
  described in Section~\ref{sec:gDInEdModel}.
\item A parameter $p$, which describes the probability of any new edge
  to be a peer-to-peer (double sided) edge.
\end{enumerate}
Setting the number of geographic regions to $l=1$ causes \gdtang\
to use the basic \dined\ model and similarly, setting the locality
parameter to $\loc=0$ approximates the \dined\ model.

For the regional distribution, we used the AS per-country
distribution data, collected by the Caida project, \cite{CAIDA-geo} in
the following way: We defined 4 large geographic regions, that
include multiple countries: NAFTA (USA, Canada and Mexico),
EMEA (Europe, Middle-East and Africa), AP (Asia-Pacific: South-east
Asia and Australia) and Latin America. Each other country
formed it's own geographic region. For each region, we defined
it's frequency as the sum of the frequencies  of ASes located in the
region.
After processing the raw data, we obtained the
distribution shown in Table~\ref{tab:dis}.

\begin{table}[t]
  \centering
\begin{tabular}{|c|c|c|}
\hline
\textbf{Region \#} & \textbf{Region ID} & \textbf{Frequency} \\
\hline
1 & NAFTA & 55.45\% \\ \hline 2 & EMEA & 18.53\% \\
\hline
3 & AP    & 8.05\% \\ \hline 4 & Latin America & 2.96\% \\
\hline
5-26 & Miscellaneous & 0.09\%-0.45\% \\
\hline
\end{tabular}
\caption{Region Size Distribution.} \label{tab:dis}
\end{table}

We used \gdtang\ to generate synthetic topologies with
Internet-like parameters. In all the experiments we used
$n=15,000$ and $m=2.11$, which match the values reported in
\cite{sw04}.

\subsection{The Fraction of Symmetric Peering Arrangements}

\begin{figure}[t]
\begin{center}
 \includegraphics[width=3.25in]{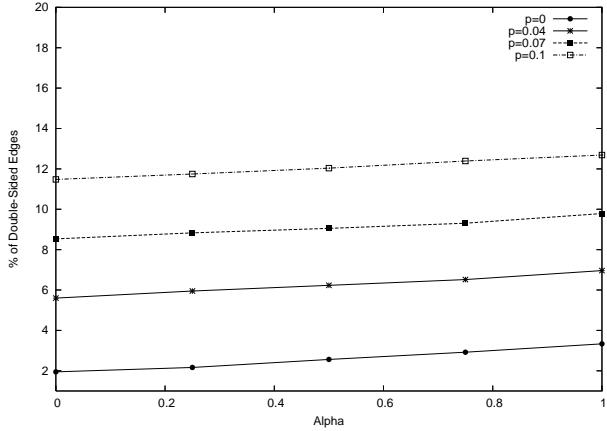}
\end{center}
\caption {Fraction of symmetric peering arrangements as a function
of locality parameter $\loc$ for various values of $p$} \label{fig:pdist}
\end{figure}

Recall that our model uses the parameter $p$, for the probability
of a peering arrangement to be symmetric. However, even when
$p=0$, the model has some probability of producing anti-parallel
edges. Therefore, to best match reality, we need to calibrate the
parameter $p$ so that total number of symmetric peering
arrangements is realistic. Gao \cite{g01} shows that about 8\% of
the peering arrangements have a symmetric peer-to-peer  nature.
Fig. \ref{fig:pdist} shows the fraction of peer-to-peer edges as
function of the locality parameter $\loc$ for $p=0,0.04,0.07,0.1$.
The figure shows that our model naturally produces 2-3\% symmetric
edges, and that the effect of the $p$ parameter is roughly
additive. So with $p=0.07$ the model produces 8.53-9.79\%
symmetric peering arrangements. All the results in the following
experiments are based on topologies produced by \gdtang\ for
$p=0.07$.

\subsection{Dense Core Analysis}

\begin{figure}[t]
\begin{center}
  \includegraphics[width=3.25in]{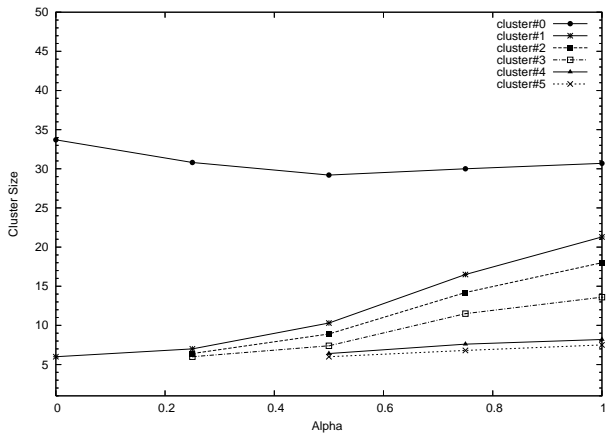}
\end{center}
\caption {Sizes of the clusters as a function of $\loc$ for $p=0.07$}
\label{fig:size}
\end{figure}

Our next experiment was designed to test the effects of the
locality parameter $\loc$. Recall that $\loc$ provably has no
effect on the degree distribution (recall
Theorem~\ref{thm:MuDIncrPowerlaw}). However, we expect $\loc$ to
have a strong effect on the clustering structure. Therefore, we
generated networks with varying values of $\loc$ and computed the
sizes of all the dense cores of over 6 nodes in each network. We
sorted the cores in decreasing order of size, from biggest to
smallest.

In order to find the Dense Core in the networks, we used the Dense
$k$-Subgraph (DkS) algorithms of \cite{fkp01,sw04}. These
algorithms search for the densest clusters (sub-graph) with a
density above a threshold: we used a value of 70\%.
Fig.~\ref{fig:size} shows the sizes of the clusters found by the algorithm
as a function of the locality parameter $\loc$. Each point on the
curve is the average over 10 random networks generated with the
same parameters.

Sagie and Wool \cite{sw04} have shown that the real AS graph has
5 dense clusters with density above 70\%. These clusters are of
sizes 43,14,8,8,7.

Fig.~\ref{fig:size} shows that a large Dense Core exists for all values of
$\loc$. However, we see that increasing $\loc$ produces
\emph{additional} cores, whose size and number grow with
$\loc$. A detailed inspection of the raw data shows that 98\% of
these secondary cores are fully contained in one of the regions,
i.e., they model the so-called Regional Cores. We believe that our
model is the first to exhibit such regional cores.

Note that the large Dense Core that our model produces is slightly
smaller that the size of 43, measured by \cite{sw04} and that
Dense Core shrinks somewhat when $\loc$ grows. The Dense Core is
not confined to a single region, so a higher locality parameter
reduces the tendency of core members to form edges with other core
members – thereby making the core less dense.

The figure shows that the \gdtang\ networks have realistic dense
and regional cores with the locality parameter $\loc$ around
$\loc=0.5$: i.e, each new edge has a 50\% probability of being a
local (regional) edge.

\subsection{Power Law Analysis} \label{sec:powerlaw}

\begin{figure}[t]
\begin{center}
\includegraphics[width=3.25in]{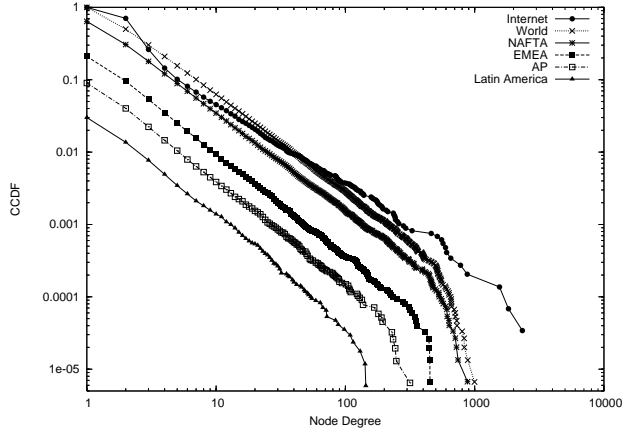}
\end{center}
\caption {The $CCDF$ of the Internet's AS-graph degree distribution,
  shown with the $CCDF$ of the synthetic networks (``world''), and
  with~4 $CCDF_R$
  curves for the largest regions we defined, on a log-log scale.}
\label{fig:ccdf}
\end{figure}

\begin{figure}[t]
\begin{center}
\includegraphics[width=3.25in]{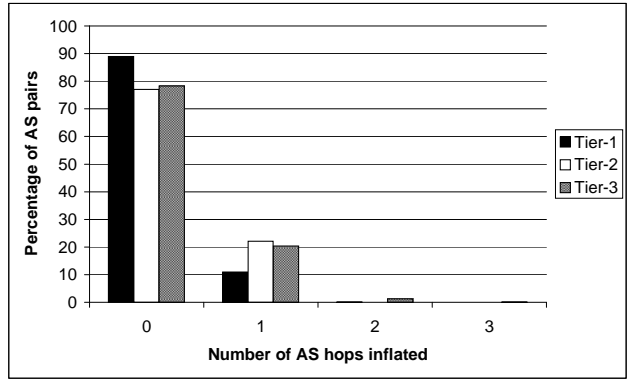} 
\end{center}
\caption {The path inflation percentage per tier in a synthetic graph,
generated with $\loc=0.5$ and $p=0.07$. } \label{fig:infl}
\end{figure}

Fig.~\ref{fig:ccdf} shows the Complementary Cumulative Density
Function for regional distribution ($CCDF_R$)%
\footnote{For any distribution of degrees in any given region $R$,
  $CCDF_R(k) = \Pr[deg_n(v) \ge k \wedge v\in R]$.  Note that if $\Pr[deg_n(v) = k]
  \propto k^{-\gamma}$ then $CCDF(k) \propto k^{-\eta} =
  k^{1-\gamma}$.}  of the degree distribution in the Internet's
AS-graph and in the \gdtang\ generated synthetic networks. For the
synthetic networks, each $CCDF_R$ curve is the average taken over
the~10 randomly generated networks.

The figure shows the well-known power-law of the Internet AS graph,
with a CCDF exponent of $\eta=1.17$.  The figure also shows that the
\gdined\ model has a fairly accurate power-law exponent of $\eta =
1.37$. Note that this is precisely the value predicted in
Theorem~\ref{thm:DIncrPowerlaw}---thus validating the estimations used
in the proofs.

The data shows that, as predicted by Theorem~\ref{thm:MuDIncrLeaf},
the model brings the
number of leaves in the network to 49\%, while the number of
leaves in the AS-graph is 30\%. Thus it seems
that the \gdined\ model produces too \emph{many} leaves.
Note, though, that the number of leaves in the AS-graph is
slightly too low for the power-law that the degree distribution
exhibits: Fig.~\ref{fig:ccdf} shows that the AS-graph's CCDF has a
``bump'' for degree values 1--4. Thus we speculate that an
additional process is taking place and affecting the frequency of
low-connectivity nodes. Exploring and modeling this phenomenon is
left for future work.

\subsection{Path inflation effects}
Gao and Wang \cite{gw02} discuss path inflation in the Internet's AS
graph due to the so-called No-Valley routing policy. They reported
that for tier-1 ISPs, 20\% of paths exhibited path inflation. For tier-2
ISPs they found 55\% path inflation and for tier-3 ISPs they found
20\% path inflation. In order to compare these findings to the behavior
on our synthetic networks, we define the No-Valley routing policy as follows:

\textsl{No-Valley Routing Policy}: an AS does not provide transit
services between any two of its providers. That is, in an AS path
($u_1,u_2....u_n$) if ($u_i,u_{i+1}$) has a provider-customer
relationship,
 then ($u_j,u_{j+1}$) must have a provider-customer relationship for any
 $i<j<n$.
 We divided the AS-es into tiers based on node degrees
 in the following way :

 \textsl{Tier1} - nodes with $Deg(node)\ge100$

 \textsl{Tier2} - nodes with $20\le{Deg(node)}<100$

 \textsl{Tier3} - nodes with $3\le{Deg(node)}<20$

 We adopted the algorithm proposed by Gao and Wang \cite{gw02} for computing
 the shortest AS path among all no-valley paths, using our definition of No-Valley
 routing policy and used it to calculate path inflation within the three tiers.
 Fig.~\ref{fig:infl} shows that the results we obtained are fairly close to those
 shown by Gao and Wang \cite{gw02}: 11\% path inflation for tier-1, 22\% path
 inflation for tier-2, and 23\% infaltion for tier-3.

\section{Conclusions and Future Work} \label{sec:Conclusions}

We have shown that our model, the \gdined\ model, significantly
improves upon previously suggested models. Most importantly, our model
produces directed graphs, which allow a much more appropriate
representation of the AS-graph's Customer-Provider peering
arrangements, as well as a representation of symmetric peer-to-peer
arrangements. Besides being more realistic,
\gdined\ even improves upon earlier, undirected, models in terms
of the (undirected) power-law exponent.
Using a simple notion of geography,
our model shows that different clustering structures can all manifest the
{\em same} power-law. Moreover, in addition to the global dense core,
for the first time, our model produces regional dense cores, when
peering arrangements have a 50\% probability of being regional.
Our model also
exhibits realistic path inflation effects. Finally, our model
is amenable to mathematical analysis, and is
implemented as a freely available network generator.

{\small

\newcommand{\etalchar}[1]{$^{#1}$}

}

\end{document}